# Dynamic Influence on Replicator Evolution for the Propagation of Competing Technologies

Elijah D. Bolluyt, Comaniciu C., *Member IEEE*

*Abstract*— This work introduces a novel modified Replicator Dynamics model, which includes external influences on the population. This framework models a realistic market into which companies, the external dynamic influences, invest resources in order to bolster their product's standing and increase their market share. The dynamic influences change in each time step of the game, and directly modify the payoff matrix of the population's interactions. The model can learn from real data how each influence affects the market, and can be used to simulate and predict the outcome of a real system. We specifically analyze how a new technology can compete and attempt to unseat an entrenched technology as the market leader. We establish a relationship between the external influences and the population payoff matrix and show how the system can be implemented to predict outcomes in a real market by simulating the rise of the Android mobile operating system over its primary competition, the iPhone, from 2009-2017.

## I. Introduction

Every real environment is in a state of perpetual flux, especially those that share a close relationship with technology. As technologies evolve in a free market, competition inevitably arises; sometimes, one company dominates, and other companies expend resources to make inroads. No matter the specifics, competition leads to each individual in the market choosing which technology to use based on the relative value of each to the user.

Replicator dynamics provides a useful framework in which to analyze competitive markets, as it models a population of identical individuals who choose a strategy based on its projected payoff, or the value to the user [1]. However, the original replicator equation's payoff is a function only of the portion of the population which employs each strategy; the strategy with the larger user base dominates [1]. In a free and active market, however, each rational company will invest assets to increase its product's adoption rate, and thus its profits.

Previous extensions of the replicator dynamic model have included feedback loops between the environment and the strategy payoffs [2], as well as the effects of localization factors on individualization of strategies [3][10]; however, while including some environmental dynamism, these approaches still do not model an environment in which external parties actively influence the market over time, while the individuals in the population remain homogeneous. Additionally, previous applications of the replicator framework to realistic scenarios have not focused on dynamic modification over time to the payoff matrix [9][11][12][13]. In addition, to the best of our knowledge, no work exists in integrating real data into such a model in order to make predictions on a real world system. These characteristics are necessary to create a model that can accurately emulate and predict a realistic marketplace.

This work introduces a general method to realistically model the dynamic influence of external factors on an unmodified population, and investigates the actions necessary for such an outside influence to achieve specific goals. We then apply this method to specifically address the case of a market that is initially dominated by an entrenched technology, describing the conditions necessary for a smaller, emerging technology to compete and, eventually, overtake the market leader. Our model not only predicts market conditions based on input information, but it also provides the user with information about the intrinsic properties of the marketplace.

Because our approach emphasizes realism and applications for modeling real markets, we conduct a simulation of the real case of the world's smartphone market from 2009-2017 to prove the efficacy of our model. This range captures the rise of the Android smartphone operating system (OS) and its ascension to dominance over the Apple iPhone, which had hitherto led the market [4].

## II. Model

### A. General Replicator Dynamics Model

Replicator dynamics models the evolution of a population of identical individuals who employ one of *n* strategies and receive a payoff from each interaction with other individuals based only on each individual's choice of strategy [1]. A player employing strategy *i* who interacts with one employing a strategy *j* receives a payoff $A_{ij}$; the set of payoffs can be expressed as a matrix:

$$A = \begin{bmatrix} A_{11} & A_{12} & \cdots & A_{1n} \\ A_{21} & A_{22} & \cdots & A_{2n} \\ \vdots & \vdots & \ddots & \vdots \\ A_{n1} & A_{n2} & \cdots & A_{nn} \end{bmatrix} \quad (1)$$

A fraction $x_i \in [0,1]$ of the population employs the strategy $i$. The replicator equation expresses the growth rate of this portion of the population:

$$\dot{x}_i = x_i((Ax)_i - x^T A x) \quad (2)$$

A strategy's usage only increases if its payoff is higher than the average payoff in the population. Thus, the expected payoff of each strategy directly modifies the individuals' choices of which strategy to employ.

The growth expressed by the replicator equation (2) captures the change in the fractions of the population that employ each strategy over time, showing how a higher payoff causes a strategy to gain prominence in the population.

Note that for the population's strategy distribution to change significantly, the elements of $A$ must be very different. It can be seen from (2) that if all elements of $A$ are approximately equal, then the payoff of each strategy will be very close to the average, and thus no strategy's growth rate will be significant.

### B. Two Strategy Replicator Dynamics Model

To model a technology market with one entrenched, dominant technology and one new technology, we formulate a population with two strategies available to each individual, where each strategy represents the usage of one of the two technologies. The payoff matrix $A$ becomes:

$$A = \begin{bmatrix} A_{11} & A_{12} \\ A_{21} & A_{22} \end{bmatrix} \quad (3)$$

*1) Equilibrium Conditions*

The two strategy replicator dynamics model has three equilibrium points at which the growth of each strategy is zero:

$$\dot{x}_1 = \dot{x}_2 = 0 \quad (4)$$

Two are trivial, wherein one strategy grows to engulf the entire market and the other is driven to extinction. The third equilibrium point can be found by using (2) and (4):

$$\hat{x}_1 = \frac{A_{22} - A_{12}}{(A_{11} + A_{22}) - (A_{21} + A_{12})} \quad (5)$$

It is important to note that the equilibrium conditions reached are dependent not only on the system's parameters, but also on the initial conditions; if one strategy starts with a market share of zero, then it never grows and the system stays constant.

*2) Impact of Payoff Matrix Element Values on System Evolution*

The elements of the payoff matrix directly influence the outcome of the market. Thus, the behavior of the system can be changed by modifying the elements of the payoff matrix in specific ways. An external party looking to influence the market should attempt to make its strategy (i.e. technology) constantly grow and finally reach equilibrium at a dominant position.

In order to grow, a strategy's payoff must be greater than the average; (2) indicates that this condition is necessary for the growth rate to be positive. This constraint leads to condition (6)

$$\dot{x}_1 > 0 \rightarrow A_{11}x_1 + A_{12}x_2 > A_{21}x_1 + A_{22}x_2 \quad (6)$$

When combined with (5), we see that growth of $x_1$ requires:

$$x_1 > \hat{x}_1 \quad (7)$$

This condition implies that the system moves away from this mixed equilibrium, not towards it, making this point an unstable equilibrium point; it also tells us that the payoff matrix $A$ must favor a low value mixed equilibrium in order for $x_1$ to continue to grow.

Condition (7), combined with (2) and (6), allow us to analyze the effects and desirable traits of each of the four payoff elements from the standpoint of strategy 1's proponent, the external influence working in favor of strategy 1.

To attain growth, $A_{11}$ should be increased according to (2). This compensates for the small initial value of $x_1$. As $x_1$ approaches dominance, $A_{11}$ can be allowed to decrease as lower growth is expected. $A_{12}$ should be kept large in the beginning to allow for the dominance of strategy 1, but its value becomes less important as $x_1$ increases. $A_{21}$ should also be kept large enough to allow for dominance of $x_1$, but it should decrease as $x_1$ increases as this will allow the growth rate of $x_1$ to increase.

Note that $A_{22}$ is not under the control of an influence of strategy 1. A technology's interactions with itself should be controlled only by the company that owns it. We can expect, however, that in a population where $x_1$ is increasing, $A_{22}$ is likely to be small in magnitude.

### C. Dynamic Influence Extension

We now formulate an extension to the payoff matrix that allows such modifications, as outlined above, to be made by an external influence on the market. We construct a column vector where each element $y_i, i \in [1, n_y]$ represents an influencing factor. These factors are inputs that affect the market and can be modified by the policies of companies in the market, such as product pricing or investments (which directly impact the quality of the product). The number of the inputs $n_y$ needed to accurately characterize the markets will depend entirely on the properties of the markets under investigation, and has no direct relation to $n$. The system will be able to model the market evolution only if the major relevant inputs that characterize that particular market are identified.

Considering $n_y$ influencing input factors, each element of the payoff matrix becomes a linear combination of these factors. The linear coefficients are expressed as a newly defined $n^2 \times n_y$ matrix $\alpha$. To illustrate the influence of the input parameters on the elements of matrix A, the payoff matrix $A$ can be re-indexed with a single index $k \in [1, n^2]$, traversing row by row, where $k = n(i-1) + j$ and $i,j$ are the original indices $i,j \in [1,n]$. Consequently, each element of the payoff matrix can be expressed as a linear combination of the input factors:

$$A_k = \alpha_k y \quad (8)$$

where $\alpha_k$ is the kth row of the coefficient matrix $\alpha$.

This representation allows the external influences to have a direct impact on the outcome of the market. As stated above, a few constraints must be placed on this matrix $\alpha$ to maintain realism; specifically, any factor controlled by company $i$ must only affect terms involving strategy $i$.

This modification makes $A$ entirely dynamic, able to take on different values during each iteration of the simulation. This allows the model to take into account the influence of external parties which change their strategies based on the market. An external party can evaluate how its input affects the payoff matrix and thus the market, and react by changing its input values accordingly. These factors may include investments a company makes into its product, pricing of its product, or any other values that have a direct impact on the quality and availability of its technology.

The value of $\alpha$ is a constant, and it represents an intrinsic property of the market's interactions with the input factors. For any particular market and input set $y$, $\alpha$ must be learned using real data. The value of $\alpha$ gives the user insight into the market, showing the influence of each input on the utility of each product. This stands in contrast to other learning methods, in which the learned parameters give no insight into the simulated system.

In our simulation, we employ a simple exhaustive search and evaluate candidate results via a least mean squares error metric, while withholding 20% of the available data in accordance with common practice [5]. When training the system, learning the parameters of $A$, this last 20% of data was withheld from the set; this tests the model's predictive capability. A mean squared error metric was used as it is computationally simple and accurately represents the validity of the result.

$$J = \frac{1}{t_f}\sum_{t=1}^{t_f}\left(x_1^*(t) - x_1(t)\right)^2 \quad (9)$$

Where $x^*(t)$ is the prediction at time $t$ and $x(t)$ is the real data at time $t$.

Due to the time dynamic nature of these modifications, the system no longer has a single guaranteed equilibrium value to which it will converge. At each time step, the system has a target equilibrium to which it would converge if the system inputs $y$ were to stay constant. The system does not necessarily reach this equilibrium, however, because the system changes over time. The system converges to a final equilibrium only when the inputs $y$, and thus the payoff matrix, converge to a constant value.

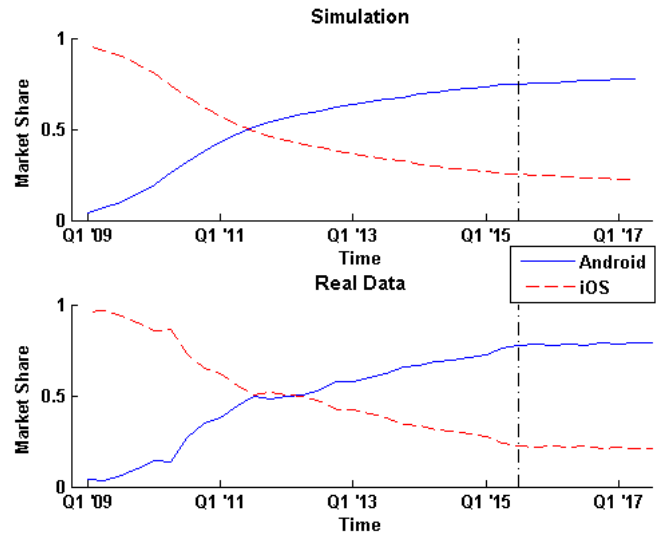

Fig 2. System outputs. The inflection point in the market and the values in the validation points are accurately predicted by the system.

## III. SIMULATION

### A. Setup and Data

To validate the accuracy of the proposed model for practical applications, a simulation was conducted on data from the global smartphone market from Quarter 1 (Q1) 2009 to Q1 2017. This range was chosen as it captures the rise of the Google Android platform to dominance over the Apple iPhone in market share. In order to simulate a two-strategy system, the smartphone market is simplified to just Android and iOS; while other platforms existed, these two grew to dominate the market as the modern wave of touchscreen smartphones developed [4].

The system takes four input factors $y_i$: average pricing for

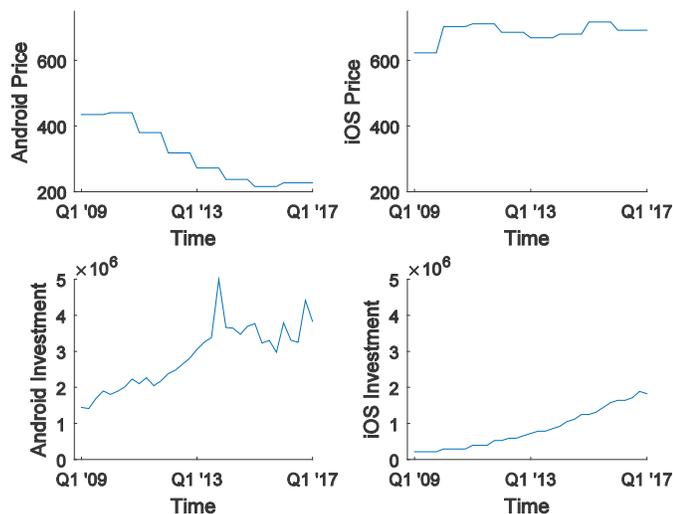

Fig 1. System inputs. These inputs contribute directly to the growth or shrinkage of each technology in the market. Note the difference in magnitudes between the investment factors, as well as the opposing trend directions of the products' pricing.

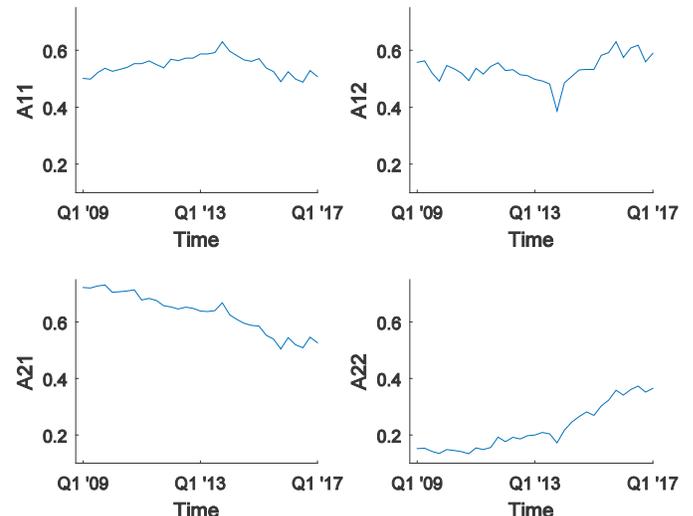

Fig 3. Payoff Matrix Coefficients. These four computed factors determine the market behavior as discussed in Section II, and their values are a direct result of the system inputs. Note the small magnitude of $A_{22}$ as well as the trend directions of $A_{11}$, $A_{12}$, and $A_{21}$.

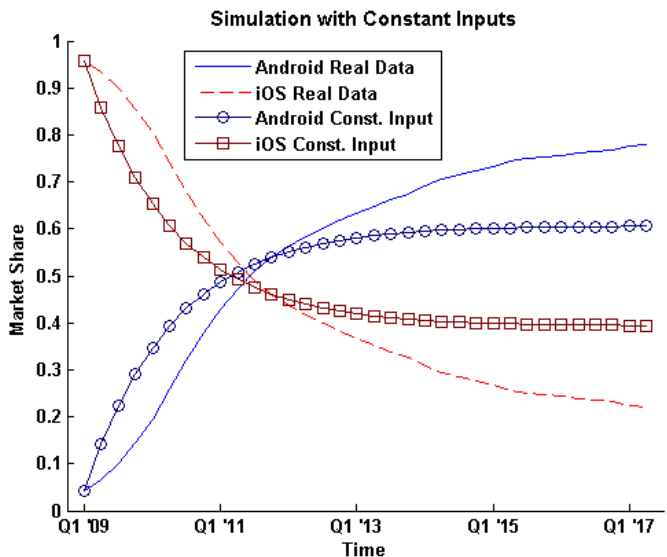

Fig 4. System outputs in simulation with constant inputs, using $\alpha$ optimized with real data. This shows that the strategy Android employed to gain dominance in the market had a great effect on increasing its final market share.

Android and iOS smartphones sold each quarter [8], and quarterly research investments in mobile from Apple [6] and Samsung [7], the largest Android phone manufacturer. These factors were chosen as they are determined by the device manufacturers' direct influence and have a direct impact on the devices' market performance. The values of these factors over the simulation period are shown in Fig. 1; these values were obtained directly from real data [6][7][8]. The data was sampled quarterly, to match the available frequency of investment data from the companies' quarterly reports. This gave a total of 33 time samples in the simulation period. Even with this relatively sparse data, the model remained effective; results are discussed further in Section III-D.

The system is initialized at the market shares determined by inputting the first real data point into the system; this is vital to obtaining a result that mirrors the reality of the market. The initial conditions are central to the determination of the system's behavior, as mentioned in Section II-B, so they must be based on the real data.

As discussed in IIC, a subset consisting of approximately 20% of the latest data, or 7 time samples, is withheld from the training set to serve as a validation set in order to determine the system's predictive capability.

### B. Implementation

A grid search is employed to optimize $\alpha$. For each candidate value, the complete evolution of the system over the training interval is computed. For each time step in the evolution, the matrix A is computed as defined in (8) and normalized, and the growth rates of the two technologies (Android and iOS) are computed (2). The market values at all time-steps are computed using the A and growth rate obtained for the candidate $\alpha$ value, and the error J between this prediction and the real data is computed. The value of $\alpha$ with the lowest error over the training interval is chosen. This $\alpha$ is used for validation in section D.

TABLE I
OPTIMIZED VALUES OF $\alpha$

| | | $y_1$ | $y_2$ | $y_3$ | $y_4$ |
|---|---|---|---|---|---|
| Simulation | $A_{11}$ | $\alpha_{11}$ | $\alpha_{12}$ | $\alpha_{13}$ | $\alpha_{14}$ |
| | $A_{12}$ | $\alpha_{21}$ | $\alpha_{22}$ | $\alpha_{23}$ | $\alpha_{24}$ |
| | $A_{21}$ | $\alpha_{31}$ | $\alpha_{32}$ | $\alpha_{33}$ | $\alpha_{34}$ |
| | $A_{22}$ | $\alpha_{41}$ | $\alpha_{42}$ | $\alpha_{43}$ | $\alpha_{44}$ |
| Real Data | | 4 | 0 | -1 | 0 |
| | | 0 | 3 | 1 | 3 |
| | | 3 | 0 | 3 | 1 |
| | | 0 | 4 | 0 | -1 |
| Constant Market (iOS dominant) | | 4 | 0 | 2 | 0 |
| | | 0 | 3 | -3 | 4 |
| | | 3 | 0 | 4 | -3 |
| | | 0 | 4 | 0 | 2 |

The value structure shows to which input $y_i$ and payoff matrix element $A_{ij}$ each element of $\alpha$ belongs, as each is a linear coefficient for one input in one payoff element.

$y_1$: Android Investment   $y_3$: Android Price
$y_2$: iOS Investment   $y_4$: iOS Price

### C. Conditions on $\alpha$ for Optimization

A is normalized to keep all elements $A_{ij} \in [0,1]$, in accordance with the standard replicator equation conditions (1). Thus the magnitude of $\alpha$ does not matter, as the system's behavior is determined by the relative size of its elements. The grid search was conducted over an interval $[-r, r]$ by integer steps; the search domain was optimized as a hyper parameter of the system over multiple trials, and a value of $r = 4$ was chosen for this simulation. The complexity of the search increases on the order of $(2r + 1)^{n_y*n^2/2}$, or $(2r + 1)^8$ in this particular simulation; this makes increasing the search interval very computationally expensive. The value of $r$ determines the resolution of the resulting $\alpha$, and must be chosen by trial and error. For our simulation we started with $r = 1$ and increased $r$ until the error between the simulated curve and the real market evolution was less than $4 \times 10^{-5}$. Since the search is conducted by integer steps and the resulting $A$ is normalized, an interval of $[-r, r]$ is equivalent to searching on an interval of $[-1,1]$ with a step size of $1/r$. A higher value for $r$ is desirable for better resolution of the search, however a high value for r induces a higher computational complexity. The complexity increases with both the number of shares in the market, $n$, and the number of input factors $n_y$. This model is well suited to technological markets, where typically only a few companies have the capability to compete effectively. Technological markets also often have a high barrier to entry, making the study of the conflct between a dominant product and a new competitor especially relevant.

Imposing practical, common sense constraints, the effect of the companies' influences on their respective products must be identical; a change in an input parameter for product 1 must cause a variation in the payoff coefficients symmetric to the result of the same change in the same input parameter for product 2. If parameters $y_1$ and $y_2$ are the same input for the two respective products, then the rates of change of $A$'s elements must be constrained by

$$\frac{\partial A_{ij}}{\partial y_i} = \frac{\partial A_{ji}}{\partial y_j}; i,j = \{1,2\} \quad (10)$$

Since these partial derivatives are linear, this can be accomplished by constraining the values of the elements of $\alpha$ which correspond to these enforced relations to be equal because the elements of $\alpha$ are the rates of change of $A$'s elements with respect to the input parameters $y$. These constraints ensure that the system remains symmetric. If we linearly index $A$ with a single variable $l$, traversing row by row as stated in Section IIC, $\frac{\partial A_l}{\partial y_k}$ corresponds to $\alpha_{lk}$ as each row of $\alpha$ corresponds to one element of $A$, as expressed in (8).

Note that any influence on technology $i$ should not exert an influence on $A_{jj}$, as discussed in Section IIB. Thus the elements of $\alpha$ corresponding to these factors are zero. In a two strategy population, there will be one such factor for each input as each input corresponds exclusively to one technology.

### D. Analysis

The simulated and real data can be seen in Fig. 2. The model was trained using the training set, and then the system was allowed to evolve on its own, employing its learned value of α and the real market input data (prices and investments) for the duration corresponding to the validation set. Its computed market share values had a mean squared error of $3.0 \times 10^{-5}$ with respect to the real data for this period. This reflects the ability of the system to predict future data with high accuracy. While this simulation approached an equilibrium state, this is due to the inputs remaining fairly constant near the end of the simulation period; such a convergence is not guaranteed in this model, as discussed in Section II-C.

The values of $A$'s elements (Fig. 3) in the simulation conformed well to expected behaviors from Section IIB. $A_{11}$ increases while technology 1 (Android) holds a small market share, and starts to decrease after the inflection point in the market. $A_{12}$ remains large with some fluctuation throughout the simulation. $A_{21}$ starts large to allow for initial growth, then decreases as Android reaches a dominant market position. $A_{22}$ is small in magnitude in comparison to the other coefficients; this is to be expected in a market where technology 1 is growing its market share. However, $A_{22}$ did increase over the simulation period, reflecting Apple's increased investment in iOS and their pricing decreases near the end of the observed period, likely an effort to grow iOS's market share.

The system's accuracy to both the real market values and the expected internal behavior indicate its efficacy in modelling the marketplace. The small input data size of 33 time steps coupled with these results indicate that the system is able to perform with much less data than would be required for many learning based models. Its replicator foundation specifically models strategy fitness as a function of population, an important feature when dealing with technologies that require a user base to stay viable. Versus a general learning method, the model only has to learn the specific nature of a particular market rather than learning the nature of markets in general; this allows it to perform well with less data, as less information has to be learned.

### E. Analysis and Significance of α

The value of $\alpha$ obtained from the simulation using real data is shown in Table 1. The values of its individual elements show the effect of each input on the payoff matrix, and thus the market. As can be expected, higher investment of each company in its own product positively effects its own product's utility ($\alpha_{11}$ and $\alpha_{42}$), and higher pricing of each product negatively effects its utility ($\alpha_{13}$ and $\alpha_{44}$). Similarly, increased research spending also increases the utility that the product's users obtain from interacting with the competitor technology ($\alpha_{22}$ and $\alpha_{31}$), indicating efforts to increase cross compatibility to increase market uptake.

A second simulation was run, in which the optimized value of α was used to predict a market outcome where the companies kept a constant level of investment and pricing, equal to that in the first time step. This was done to assess the effectiveness of the strategies employed by each company, and determine what effect the feedback from the market had on the outcome. The results of this simulation can be seen in Fig. 4; Android still becomes dominant, but its market share saturates at a lower value. This would seem to indicate that Android's strategy of decreasing price and increasing investment payed off significantly.

Finally, a third simulation was conducted in which the market shares were held constant. This was done to observe what type of market would have allowed iOS to remain dominant, given the real input data. The value of α optimized for this hypothetical situation is also shown in Table 1. Investments play a similar role in increasing a product's utility, but higher prices now have a large positive effect on the utility of each product ($\alpha_{13}$, $\alpha_{44}$). Such an unrealistic property makes sense because iOS devices have a much higher price compared to Android devices throughout the simulation period. If consumers wanted to pay a higher price, then this would have improved iOS's market share size rather than shrinking it.

Analyzing these simulations shows that the properties of α reflect real properties of the market. This direct relation is the core advantage of this model over other learning based methods: the learned parameters give direct insight into the market reaction relative to the inputs. This could potentially allow a market player to not only simulate strategies and observe the outcome, but also to understand how the individual inputs under its control could modify the market share evolution. The players could then devise a strategy that achieves the desired outcome based on this knowledge of the market.

### IV. CONCLUSION

This work extends the Replicator Dynamics framework to analyze a system with external influencing factors. It modifies the payoff matrix of the core system to adapt to external factors, and learns parameters to model that dependence. This enables the system to contend with external factors that influence an otherwise symmetric population, an ability not present in previous replicator formulations. This modified framework was applied to the rise of the Android platform to mobile OS market dominance between 2009 and 2017; the proposed system accurately modeled the evolution of each technology's market share for several fiscal quarters, employing parameters obtained via optimization over a training set. The simulation also produced intermediate parameter behavior which closely corresponded with theoretical predictions for a scenario such as the rise of Android over iOS, starting from a small initial market share. This simulation result indicates that the system can be employed to accurately model and predict the evolution of markets in real-world scenarios. This model could aid in predicting the outcome of variations in strategy of companies in a marketplace, making it a useful tool to create new strategies which can produce a desired outcome.

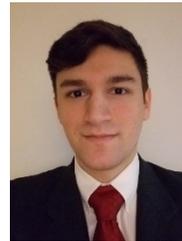

**Elijah D. Bolluyt** received the B.S. degree in engineering physics and the M.E. degree in electrical engineering from Stevens Institute of Technology, Hoboken, NJ, in 2017. He is currently pursuing the Ph.D. degree in computer engineering at Stevens Institute of Technology, Hoboken, NJ.

During the summer of 2015, he was a Scholars summer researcher with the Department of Physics and Engineering Physics at Stevens Institute of Technology. During the summer of 2016, he was an Engineering Intern at Kulite Semiconductor, Inc. in Leonia, NJ. His current research interests focus on the development of computational frameworks, employing game theory and machine learning techniques.

Mr. Bolluyt is a recipient of the Provost's Doctoral Fellowship (Stevens Institute of Technology).

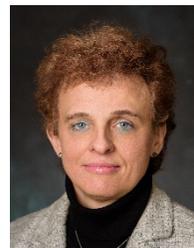

**Cristina Comaniciu** received the M.S. degree in electronics from the Polytechnic University of Bucharest in 1993, and the Ph.D. degree in electrical and computer engineering from Rutgers University in 2002.

From 2002 to 2003 she was a postdoctoral fellow with the Department of Electrical Engineering, Princeton University. Since August 2003, she is with Stevens Institute of Technology, Department of Electrical and Computer Engineering, where she is now an Associate Professor and serves as Associate Department Chair for Graduate Studies. In Fall 2011 she was a visiting faculty fellow with the Department of Electrical Engineering, Princeton University. She served as an associate editor for the IEEE COMMUNICATION LETTERS (2007-2011).

Professor Comaniciu is a recipient of the 2007 IEEE Marconi Best Paper Prize Award in Wireless Communications and of the 2012 Rutgers School of Engineering Distinguished Young Alumnus Medal of Excellence. She is a coauthor of the book Wireless Networks: Multiuser Detection in Cross-Layer Design (Springer, NY). Her research interests are focused on applications of game theory, evolutionary games and machine learning for resource management and optimization of complex distributed networks.